\pdfoutput=1

\documentclass[11pt]{article}

\usepackage{ACL2023}
\usepackage{times}
\usepackage{latexsym}
\usepackage[T1]{fontenc}
\usepackage[utf8]{inputenc}
\usepackage{microtype}
\usepackage{inconsolata}
\usepackage{enumitem}
\usepackage{tikz}
\usepackage{amsmath}
\usepackage{booktabs}
\usepackage{lipsum}
\usepackage{listings}
\usepackage{epsfig,endnotes}
\usepackage{graphicx}
\usepackage{array, multirow}
\usepackage{makecell}
\usepackage{tabularx}
\usepackage{xspace}
\usepackage{array}
\usepackage{caption}
\usepackage{glossaries}
\usepackage{varwidth}
\usepackage{xcolor}
\usepackage{colortbl}
\usepackage{subcaption}
\usepackage{mathtools} 
\usepackage{amsfonts} 
\usepackage{verbatim}
\usepackage{framed}
\usepackage{inconsolata}
\usepackage{fancyvrb}
\usepackage{amsthm}
\usepackage{dblfloatfix}
\usepackage[english]{babel}
\usepackage[most]{tcolorbox}

\usepackage{xspace}
\usepackage{color}

\newcommand{\tool}{LDB\xspace}
\newcommand{\para}[1]{\paragraph{#1}\quad}
\lstdefinestyle{demo}{
    basicstyle=\fontsize{9}{9}\ttfamily,
    keywordstyle=\color{blue},
    commentstyle=\color{gray},
    stringstyle=\color{green},
    showstringspaces=false,
    breaklines=true,
    breakatwhitespace=false,
    breakindent=0pt,
    escapeinside={(*@}{@*)},
    literate={á}{{\'a}}1 {ã}{{\~a}}1 {é}{{\'e}}1,
}

\definecolor{cadmiumgreen}{rgb}{0.0, 0.42, 0.24}
\definecolor{forestgreen}{HTML}{009B55}

\makeatletter
\def\@fnsymbol#1{\ensuremath{\ifcase#1\or \dagger\or \ddagger\or
   \mathsection\or \mathparagraph\or \|\or **\or \dagger\dagger
   \or \ddagger\ddagger \else\@ctrerr\fi}}
\makeatother
\title{Debug like a Human: A Large Language Model Debugger via Verifying \\Runtime Execution Step by Step}

\author{
    Li Zhong\quad Zilong Wang\thanks{$\ $  Corresponding authors.}\quad Jingbo Shang\footnotemark[1]\\
    University of California, San Diego\\
    \texttt{\{lizhong, zlwang, jshang\}@ucsd.edu}
}

\begin{document}
\maketitle
\begin{abstract}

Large language models (LLMs) are leading significant progress in code generation. Beyond one-pass code generation, recent works further integrate unit tests and program verifiers into LLMs to iteratively refine the generated programs. However, these works consider the generated program as an indivisible entity, which falls short for LLMs in debugging the programs, especially when the programs contain complex logic flows and data operations. In contrast, when human developers debug programs, they typically set breakpoints and selectively examine \emph{runtime execution} information. The execution flow and the intermediate variables play a crucial role in the debugging process, yet they are underutilized in the existing literature on code generation. In this study, we introduce Large \underline{L}anguage Model \underline{D}e\underline{b}ugger (\tool{}), a novel debugging framework that enables LLMs to refine their generated programs with the runtime execution information. Specifically, \tool segments programs into basic blocks and tracks the values of intermediate variables after each block throughout runtime execution. This allows LLMs to concentrate on simpler code units within the overall execution flow, verify their correctness against the task description block by block, and effectively pinpoint any potential errors. Experiments demonstrate that \tool consistently enhances the baseline performance by up to 9.8\% across the HumanEval, MBPP, and TransCoder benchmarks, archiving new state-of-the-art performance in code debugging for various LLM selections.

\end{abstract}

\begin{figure}
    \centering
    \includegraphics[width=0.95\linewidth]{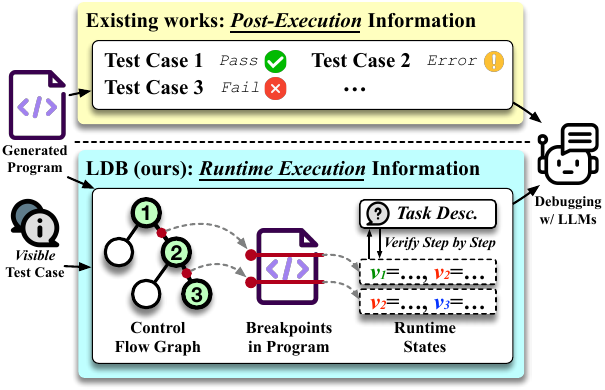}
    \caption{Comparison of \tool and existing debugging works. Existing works treat the programs as an indivisible entity and depend on the post-execution feedback for debugging, while \tool leverages the runtime execution information, tracking the values of intermediate variables and verifying basic blocks against the task description step by step.}
    \label{fig:overview}
\end{figure}

\begin{figure*}[t]
    \centering
    \includegraphics[width=0.98\textwidth]{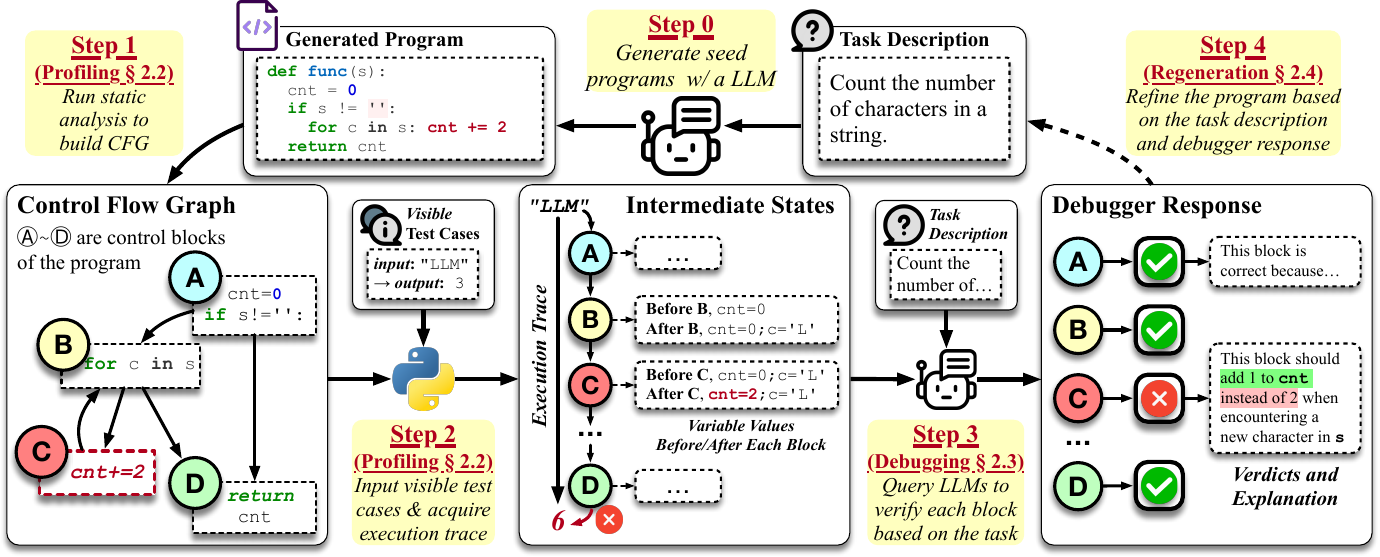}
    
    \caption{Illustration of the debugging workflow of \tool. A code generator is prompted to generate the seed programs (Step 0). 
    \textbf{Profiling (\S~\ref{subsec:profiling}):} \tool decomposes the seed program into basic blocks based on the control flow graph (Step 1), and feeds in a failed visible test case to acquire the execution trace (Step 2). 
    \textbf{Debugging (\S~\ref{subsec:debugging}):} \tool further inspects the runtime states of variables after each basic block during the runtime execution. Gathering the runtime execution information, \tool queries a LLM for verdicts on the correctness of the blocks in the relation to the task description (Step 3). 
    \textbf{Regeneration (\S~\ref{subsec:regeneration}):} Finally, the LLM regenerates a refined program with the debugging feedbacks by \tool (Step 4).}
    
    \label{fig:main-figure}
\end{figure*}

\section{Introduction}
Code generation is a critical yet challenging task that has various downstream applications, such as text-to-code generation~\cite{chen2021evaluating,yin2017syntactic,li2022competition}, code translation~\cite{roziere2020unsupervised}, and code autocompletion~\cite{li2017code, raychev2014code}. 
Recent progress in large language models (LLMs)~\cite{li2023starcoder,roziere2023code, achiam2023gpt,zhou2023language,muennighoff2023octopack} significantly boosts the performance of code generation and demonstrates a promising potential to be generally applied in different requirements and tasks~\cite{shinn2023reflexion, gu2023llm, yuan2023evaluating}. 
However, generating correct programs is not a one-time effort. Existing works suggest enhancing code generation through multiple sampling~\cite{zhang2023coder, shinn2023reflexion}, self-consistency~\cite{le2023codechain, huang2023enhancing, chen2022codet}, and candidates ranking~\cite{shi2022natural, ni2023lever, zhang2023algo}. Despite these advanced approaches, they still fall short on basic programming questions from the HumanEval and MBPP datasets. This underscores the limitations of \emph{single-pass} program generation.


Recognizing this, a series of works have been proposed to refine the programs generated in a single pass, based on feedback from either human annotator~\cite{chen2023improving, wu2023fine} or LLMs themselves~\cite{tang2023explain, chen2023teaching}. This refinement process is akin to \emph{debugging} in programming practices by human developers. \citet{chen2023teaching,jiang2023selfevolve} introduce unit test results and error messages to LLMs. These approaches allow LLMs to reflect on potential mistakes and generate corrected programs.
Nevertheless, considering the debugging process by human developers, it is sub-optimal to solely depend on these \emph{post-execution information} to debug the program, especially in cases involving complex data structures and control flows. 
In fact, when human developers encounter a buggy program, they do more than just collect the program's outputs.
They delve into the runtime execution to observe the execution traces\footnote{Refer to Appendix~\ref{app:pl} for detailed introduction of execution trace, control flow graph, and basic block.\label{foot1}} and examine the intermediate variables by setting breakpoints.
When the intermediate execution states deviate from their intention, developers pinpoint the bugs and make the corrections. This is a common workflow for well-known interactive debuggers such as GDB~\cite{stallman1988debugging} and PDB~\cite{pdb}.


To this end, we propose \tool, a \textit{large \underline{l}anguage model \underline{d}e\underline{b}ugger} that refines programs generated by LLMs using runtime execution information, emulating the debugging practices of human developers. 
As shown in Figure \ref{fig:overview}, feeding in a visible test case,
\tool segments the execution trace into basic blocks\textsuperscript{\ref{foot1}} based on the control flow graph\textsuperscript{\ref{foot1}}. \tool tracks the intermediate variables at the end of each basic block, similar to the breakpoints set by developers. 
After gathering runtime execution information, \tool queries LLMs for verdicts on each code block's correctness and explanations of the execution flow in relation to the coding task. This approach allows language models to concentrate on simpler code units, verify intermediate states against the task description, and pinpoint potential bugs. Consequently, it effectively debugs the program and improves the quality of code generation.

We validate \tool on three code generation benchmarks, including HumanEval~\cite{chen2021evaluating} and MBPP~\cite{austin2021program} for text-to-code generation, and TransCoder~\cite{roziere2020unsupervised} for code translation. 
We conduct experiments using the proprietary model, GPT-3.5~\cite{achiam2023gpt}, and the open-sourced models, StarCoder~\cite{li2023starcoder} and CodeLlama~\cite{roziere2023code}. Experiments demonstrate that \tool consistently improves code generation accuracy across various LLM backbones and achieves state-of-the-art performance in debugging programs. 
Worth mentioning, even with the programs generated by more powerful code generators, such as GPT-4~\cite{achiam2023gpt} and Reflexion~\cite{shinn2023reflexion}, \tool can still detect errors overlooked by previous advanced methods, thereby enhancing the capabilities of code generation even further.
We summarize our contribution as follows:
\begin{itemize}[leftmargin=*,nosep]
    \item We propose a large language model debugger, \tool, which takes the very first step on incorporating runtime execution information into LLMs to debug generated programs.
    \item We leverage basic blocks in the execution traces to properly segment the programs into smaller, simpler code units, allowing LLMs to verify them step by step against users' intention and effectively pinpoint the potential bugs.
    \item Extensive experiments on three code generation benchmarks demonstrate the effectiveness of \tool in debugging generated programs across different LLM backbones.
\end{itemize}
\noindent\textbf{Reproducibility.} The code will be released on Github\footnote{\url{https://github.com/FloridSleeves/LLMDebugger}}.

\section{Large Language Model Debugger}
\paragraph{Problem Formulation.} 
\label{sec:problem-formulation}

We follow the problem formulation of program debugging in \citet{chen2023teaching}.
In a code generation task, each sample can be represented as a triplet $(Q, T_v, T_h)$, where $Q$ represents the task description, $T_v$ and $T_h$ are visible and hidden test cases. 
In the text-to-code generation task, $Q$ consists of a brief paragraph in natural language that outlines the intended goal of the task. In the code translation task, $Q$ represents a program written in a language other than the target language, and the objective is to translate this program into the target language. 
A code generator is first provided with $Q$ and $T_v$ to generate the seed program $A_0$.
In the debugging stage, a debugger further refines $A_0$ and finally outputs a new program $A^\ast$. $A^\ast$ will be tested using the hidden tests $T_h$ to evaluate whether it is correct or not. $T_h$ is only used in the evaluation stage and not visible during the code generation or debugging stages.

\subsection{Overview}
We show the workflow of \tool in Figure~\ref{fig:main-figure}. We prompt a LLM to generate seed programs. This serves as the starting point of debugging if they fail any of the visible test cases. During debugging, \tool feeds in the failed visible test case to the seed program and collects the runtime execution information, including the execution trace and the runtime variable values after each basic block (\textbf{Section~\ref{subsec:profiling}}). Then, \tool queries a LLM to verify the correctness of basic blocks step by step via comparing them with the task description (\textbf{Section~\ref{subsec:debugging}}). Considering the block-wise runtime execution and the task description, \tool queries the LLM again to detect buggy blocks and regenerate the program. \tool repeats these three steps iteratively until the new solution pass all the visible tests, or the maximum debugging iteration is reached (\textbf{Section~\ref{subsec:regeneration}}).

\subsection{Profiling}
\label{subsec:profiling}
In the step  of profiling, \tool collects the runtime execution information when running the generated program over the failed visible test cases. It collects the execution trace and then segments the trace into basic blocks to inspect intermediate variables.

\paragraph{Execution Traces.} In static analysis, each program corresponds to a unique control flow graph (CFG) where each node in the graph is a code basic block, as shown in Figure \ref{fig:main-figure}. Each basic block is a straight-line sequence of code with only one entry point and one exit. The CFG represents all paths might be traversed through a program during its execution. After \tool feeding in a visible test case, the control flow goes through a sequence of basic blocks and the path is denoted as the execution trace, $[B_1, B_2, \ldots, B_n]$, where $B_i$ is a basic block in the trace and $n$ is the length of the trace.


\paragraph{Intermediate States.}  Given an execution trace $[B_1, B_2, B_3, \ldots, B_n]$, we execute the first $i$ blocks and collect all the variables in the scope along with their runtime values. We denote the state set as $V_i = \{ v = \hat{v} | v \in \bigcup B_{\le i}\}$, where $v$ is a variable used in the first $i$ blocks and $\hat{v}$ is its runtime value after the $i$-th block. We define the intermediate state after the first $i$ blocks as $(V_{i-1}, B_i, V_i)$, where $V_{i-1}$ represents the entry states of the block, $B_i$ is the current code block to execute, and $V_i$ provides the actual execution results after $B_{i}$.



\subsection{Debugging}
\label{subsec:debugging}
The block-wise intermediate states determined by profiling provide a comprehensive illustration for the runtime execution. In Debugging, we integrate the intermediate states into prompts and query a LLM to verify whether the basic blocks align with the intended semantics in the task description $Q$.

\paragraph{Debugging Verdicts.} For each intermediate state throughout the trace, $(V_{i-1}, B_{i}, V_{i})$, the LLM is acquired to make a verdict on its correctness $D_i\in \{\texttt{True}, \texttt{False}\}$, and elaborate the explanation $E_i$. If the LLM detects any buggy code block, \tool includes the message in the debugging response.

\paragraph{Selective Debugging.} Loops and recursion are common in programming, potentially leading to extensive execution traces. If we include the lengthy traces directly in the prompt, it is highly likely to exceed the maximum token limit of LLMs. This is similar to what occurs with humans in program development. When the execution is lengthy, developers may only examine a few blocks and skip the other long and tedious execution traces when detecting bugs. Inspired by the human practice, \tool selectively samples $N_b$ blocks from program traces to ensure the total length of runtime information within the max token limit of LLMs.


\paragraph{Batch Debugging} In our proposed \tool, the intermediate states following each basic block are determined during the execution of the seed program with test cases. Thus, \tool can batch these states together and query the LLM for debugging verdicts. This significantly improves the token efficiency of \tool and alleviates the pitfall of repeatedly sending lengthy context to LLMs in iterative refinement~\cite{ge2023context, hu2023tree}. Specifically, the batch debugging query process is as follows,
\begin{align*}
                         \{V_0&, B_1, V_1, B_2, ..., B_n, V_n\}\\
\xrightarrow{\text{LLM}} &\{(D_1, E_1), ..., (D_n, E_n)\}
\end{align*}
where $V_i$ is the set of variables and their runtime values after the $i$-th blocks, $B_i$ is the $i$-th block in the trace, $D_i$ is the debugging verdict from the LLM, and $E_i$ is the corresponding explanation.




\subsection{Regeneration}
\label{subsec:regeneration}
The runtime execution information helps accurately localize buggy code blocks, allowing LLMs to concentrate specifically on these areas during the regeneration process.
In Regeneration, \tool collects the debugging verdicts $D$ and explanations $E$, and incorporate them along with the task description $Q$ into the prompt. Then, \tool queries the LLM again to generate the refined program.
\tool iteratively runs \textit{Profiling}, \textit{Debugging}, and \textit{Regeneration}, until the refined program passes all visible test cases, or the maximum debugging iteration is reached. We test the finalized solution $A^\ast$ using the hidden test cases $T_h$ to evaluate the performance.

\section{Experiments}
We evaluate \tool on three code generation benchmarks: HumanEval~\cite{chen2021evaluating}, TransCoder~\cite{roziere2020unsupervised}, and MBPP~\cite{austin2021program}. HumanEval and MBPP are for text-to-code generation, where the task description is a brief passage outlines the intended functionality of the program to be generated. TransCoder is for code translation which requires to translate a program from C++ into Python. The task description of TransCoder consists of a C++ program to be translated. We compute Pass@1 accuracy with hidden test cases for assesment.
We conduct experiments with the proprietary LLM, GPT-3.5 (\texttt{turbo-0613})~\cite{achiam2023gpt}, and the open-source LLMs, CodeLlama (\texttt{34B-Instruct})~\cite{roziere2023code} and StarCoder ($\sim$15B)~\cite{li2023starcoder} as backbones. 



\subsection{Experiment Setup}
\label{sec:setup}
We generate the seed programs following the same prompts and generation parameters used in our compared method \citet{chen2023teaching}. We set the maximum number of debugging iterations as 10. More detailed implementation details are reported in Appendix~\ref{app:param}.
To obtain visible test cases for HumanEval, we extract the given visible test cases from the task description. For MBPP, we use the first test case of each problem as the visible test case and use the rest as hidden test cases.
For TransCoder, we include all test cases from the dataset as visible test cases\footnote{All test cases can be generated by running the original C++ programs, which are visible to LLMs.}. The experiment settings on MBPP and TransCoder are the same as the prior works~\cite{chen2023teaching,shi2022natural,ni2023lever}.
After finalizing the code solution, we compute the Pass@1 accuracy with hidden test cases to evaluate the performance.






\begin{table*}[!htb]
  \centering
  \small
  \resizebox{\linewidth}{!}{
    \setlength{\tabcolsep}{2.2mm}{
    \begin{tabular}{llcccccc}
    \toprule
    \multirow{3}[6]{*}{\textbf{Model (\# Param.)}} & \multirow{3}[6]{*}{\textbf{Debugger}} & \multicolumn{6}{c}{\textbf{Dataset}} \\
\cmidrule{3-8}          &       & \multicolumn{2}{c}{\textbf{HumanEval}} & \multicolumn{2}{c}{\textbf{TransCoder}} & \multicolumn{2}{c}{\textbf{MBPP}} \\
\cmidrule(lr){3-4} \cmidrule(lr){5-6} \cmidrule(lr){7-8}         &       & \textbf{Acc. $\uparrow$} & $\Delta \uparrow$ & \textbf{Acc. $\uparrow$} & $\Delta \uparrow$ & \textbf{Acc. $\uparrow$} & $\Delta \uparrow$ \\
    \midrule
    \multirow{4}[2]{*}{GPT-3.5 ($\ge$175B$^\dagger$)} 
          & Baseline (w/o debugger)                            & 73.8  &    & 82.3  &     & 67.6  &    \\
          & SD (+Expl.)~\cite{chen2023teaching} & 81.1  & \textcolor{cadmiumgreen}{+7.3}  & 85.9  & \textcolor{cadmiumgreen}{+3.6}  & 74.4  & \textcolor{cadmiumgreen}{+6.8} \\
          & SD (+Trace)~\cite{chen2023teaching} & 80.5  & \textcolor{cadmiumgreen}{+6.7}  & 86.1  & \textcolor{cadmiumgreen}{+3.8}  & 72.6  & \textcolor{cadmiumgreen}{+5.0} \\
          & \textbf{LDB (ours)}                          & \textbf{82.9}  & \textcolor{cadmiumgreen}{\textbf{+9.1}}  & \textbf{87.7}  & \textcolor{cadmiumgreen}{\textbf{+5.4}}  & \textbf{76.0}  & \textcolor{cadmiumgreen}{\textbf{+8.4}} \\
    \midrule
    \multirow{4}[1]{*}{CodeLlama (34B)} 
          & Baseline (w/o debugger)                            & 49.4  &     & 69.8  &    & 51.2  &\\
          & SD (+Expl.)~\cite{chen2023teaching} & 53.0  & \textcolor{cadmiumgreen}{+3.6}  & 79.4  & \textcolor{cadmiumgreen}{+9.6}  & 55.6  & \textcolor{cadmiumgreen}{+4.4} \\
          & SD (+Trace)~\cite{chen2023teaching} & 54.3  & \textcolor{cadmiumgreen}{+4.9}  & 76.4  & \textcolor{cadmiumgreen}{+6.6}  & 57.2  & \textcolor{cadmiumgreen}{+6.0} \\
          & \textbf{LDB (ours)}                          & \textbf{55.5}  & \textcolor{cadmiumgreen}{\textbf{+6.1}}  & \textbf{79.6}  & \textcolor{cadmiumgreen}{\textbf{+9.8}}  & \textbf{57.4}  & \textcolor{cadmiumgreen}{\textbf{+6.2}} \\
    \midrule
    \multirow{4}[1]{*}{StarCoder (15B)} 
          & Baseline (w/o debugger)                            & 39.0  &      & 61.8  &     & 51.6  &\\
          & SD (+Expl.)~\cite{chen2023teaching} & 38.4  & \textcolor{red}{-0.6}  & 68.9  & \textcolor{cadmiumgreen}{+7.1}  & 54.4  & \textcolor{cadmiumgreen}{+2.8} \\
          & SD (+Trace)~\cite{chen2023teaching} & 39.0  &                          +0.0  & 65.7  & \textcolor{cadmiumgreen}{+3.9}  & 54.8  & \textcolor{cadmiumgreen}{+3.2} \\
          & \textbf{LDB (ours)}                          & \textbf{39.6}  & \textcolor{cadmiumgreen}{\textbf{+0.6}}  & \textbf{69.8}  & \textcolor{cadmiumgreen}{\textbf{+8.0}}  & \textbf{55.4}  & \textcolor{cadmiumgreen}{\textbf{+3.8}} \\
    \bottomrule
    \end{tabular}%
    }
    }
    \caption{Results of \tool and Self-Debugging~\cite{chen2023teaching} (denoted as SD) on HumanEval, TransCoder, and MBPP with GPT-3.5, CodeLlama, and StarCoder. Accuracy is calculated based on Pass@1. The improvement (denoted as $\Delta$) is measured against the baseline (w/o debugger). $^\dagger$ We assume the parameter number in GPT-3.5 is larger than that of GPT-3 (175B).}
  \label{tab:main}%
\end{table*}%

\subsection{Compared Methods}
We evaluate the seed programs and label the performance as \textbf{Baseline (w/o debugger)}. We compare \tool against two \textit{rubber duck debugging} methods from \citet{chen2023teaching}: 
\textbf{Self-Debugging (+Expl.)} which prompts LLMs to explain generated programs line-by-line as feedback, and 
\textbf{Self-Debugging (+Trace)} which prompts LLMs to dry run generated programs as feedback.

We reproduce the Self-Debugging methods following the instructions in \citet{chen2023teaching} due to the unavailability of open-source code.
When referring to ``Self-Debugging'', we default to the method with higher accuracy among the two methods unless otherwise specified.
Throughout the evaluation process, we ensure that all debugging methods utilize the \textit{same} LLM settings, visible test cases, seed programs, and prompts formats. This ensures a fair comparison and eliminates potential disruptions caused by changes in prompt formats.



\subsection{Main Results}
We compare \tool with the baseline debugging methods on HumanEval, TransCoder, and MBPP, and present the result in Table~\ref{tab:main}. 
We observe that \tool consistently achieves improvements of up to 9.8\% on all datasets across different LLM backbones. Specifically, compared to Self-Debugging which prompts LLMs to dry run or explain the program, \tool achieves higher and more stable performance gain over the baseline by introducing the actual runtime execution information. 

We attribute the advantage of \tool to the fine-grained debugging feedback and the runtime information as external supplements to the LLM self-correction.
The detailed block-level debugging responses help LLMs concentrate on the buggy areas in the program to better align the program with the task description. Moreover, as pointed out by \citet{huang2023large}, LLMs has limited self-correct reasoning abilities. Particularly in code generation, LLMs are prone to mistakes when reflecting on program execution. This is due to their inability to accurately calculate concrete variable values and predicting execution flow at branches or loops. 
The inaccurate feedback from LLMs could misguide the program debugging and refinement, which explains why Self-Debugging (+Expl.) and Self-Debugging (+Trace) fail to improve the seed programs on HumanEval with StarCoder. 
On the contrary, \tool generates the debugging verdicts and explanations based on accurate intermediate values and execution flows, guiding the generated programs towards the correct answer.






Worth mentioning, the visible test cases are also provided to the LLM during the initial seed program generation. However, their utility in code generation is limited, as evidenced by the sub-optimal performance of Baseline (w/o debugger).
This aligns with our assumption that actual runtime execution information significantly helps LLMs ground their reasoning, thereby improving their ability to generate better semantically aligned code.

\begin{table}[t]
  \centering
    \small
  \resizebox{\linewidth}{!}{
    \setlength{\tabcolsep}{1.7mm}{
\begin{tabular}{llll}
\toprule
\multirow{2}[4]{*}{\textbf{Debugger}} & \multicolumn{3}{c}{\textbf{Code Generator}} \\
\cmidrule{2-4}      & \textbf{GPT-3.5} & \textbf{GPT-4} & \textbf{Reflexion} \\
\midrule
\textit{(w/o debugger)} & 73.8  & 87.2  & 91.5 \\
SD \scriptsize{(GPT-3.5)}    & 81.1 \scriptsize{\textcolor{cadmiumgreen}{{(+7.3)}}}
                             & 88.4 \scriptsize{\textcolor{cadmiumgreen}{{(+1.2)}}}
                             & 92.1 \scriptsize{\textcolor{cadmiumgreen}{{(+0.6)}}} \\
\textbf{LDB \scriptsize{(GPT-3.5)}}  & \textbf{82.9} \scriptsize{\textcolor{cadmiumgreen}{\textbf{(+9.1)}}} 
                                     & \textbf{89.6} \scriptsize{\textcolor{cadmiumgreen}{\textbf{(+2.4)}}} 
                                     & \textbf{95.1} \scriptsize{\textcolor{cadmiumgreen}{\textbf{(+3.6)}}} \\
\bottomrule
\end{tabular}%
    }
    }
  \caption{Results of \tool and Self-Debugging (denoted as SD) on HumanEval with seed programs from GPT-3.5, GPT-4, and Reflexion. We use GPT-3.5 as the debugging backbone. \tool can detect the subtle bugs overlooked by the powerful code generation method and improve the performance even further.}
  \label{tab:diff_seed}
\end{table}%


\subsection{Results on Advanced Code Generators}
To further demonstrate the effectiveness of \tool, we apply \tool and Self-Debugging to debug the seed programs from advanced code generators, GPT-4~\cite{achiam2023gpt} and Reflexion~\cite{shinn2023reflexion}. We conduct the analysis on HumanEval as an example. We query GPT-4 (\texttt{1106-preview}) to generate seed programs in the same setting introduced in Section~\ref{sec:setup}. As for Reflexion, we utilize the corresponding generated programs published in the official Github repository\footnote{\url{https://github.com/noahshinn/reflexion}} as the seed programs. GPT-4 and Reflexion are considered as more powerful code generators that already achieve superior performance without any debuggers. 

The results are shown in Table~\ref{tab:diff_seed}. The LLM backbone of Self-Debugging and \tool is GPT-3.5 (\texttt{turbo-0613}), which is weaker than the code generators, GPT-4 and Reflexion. We list the performance of GPT-3.5 in the table for reference. 
Despite the weaker LLM backbone, both Self-Debugging and \tool can refine the programs in the debugging process. This highlights the advantage of introducing a debugging stage in code generation with LLMs. Furthermore, \tool surpasses Self-Debugging in debugging and refining programs. It can improve performance on HumanEval even further and achieve a new state-of-the-art result (95.1\%) in code generation by debugging the seed programs from Reflexion. This indicates that \tool is able to examine the runtime execution and correct bugs overlooked by the advanced code generators, serving as an orthogonal supplement to current code generation techniques.



\begin{figure}[t]
    \centering
    \includegraphics[width=0.95\linewidth]{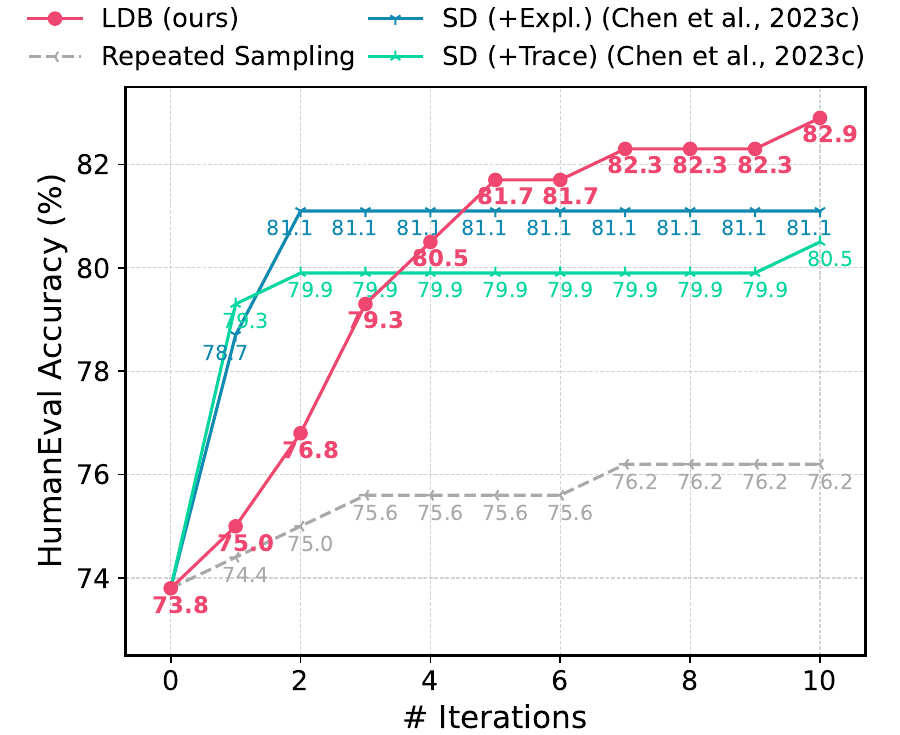}
    
    \caption{Performance at each debugging iteration on HumanEval with \tool, Repeated Sampling, and Self-Debugging using GPT-3.5 as the backbone. SD stands for Self-Debugging. \tool exhibits a continuing growth potential with the increasing iterations and achieves the best performance in debugging after 10 iterations.}
    
    \label{fig:rounds}
\end{figure}

\subsection{Performance vs. Debugging Iterations}

 In Figure~\ref{fig:rounds}, we plot the performance of \tool, Repeated Sampling, Self-Debugging (+Expl.), and Self-Debugging (+Trace) across each iteration on HumanEval using GPT-3.5. We introduce Repeated Sampling as a straightforward comparison method, where we repeatedly sample coding solutions from the program generator until the solution passes the visible test. The performance at each iteration is computed in the same way as \tool. We run these methods up to 10 iterations to examine the performance tendency. We show the performance of \tool across 20 iterations in Appendix~\ref{fig:20turns} to explore the continuous improvement trend. 

\paragraph{Continuous Debugging Potential of \tool}
In Figure~\ref{fig:rounds}, with increasing debugging or resampling rounds, all methods refine the seed program and improve the performance. Particularly, \tool continuously improves the performance across the debugging iterations and achieves the best debugging performance despite the slightly slow rising speed. In contrast, Self-Debugging nearly stops improving the performance after 2 iterations, as also observed in \citet{chen2023teaching}.

\paragraph{Necessity of Runtime Information.}
From Figure~\ref{fig:rounds}, we observe that the performance of Self-Debugging presents a similar trend to Repeated Sampling after 3 rounds. They both stops effectively improving the performance at an early stage (around 2 $\sim$ 3 iterations). This phenomenon reveals a fundamental difference between \tool and Self-Debugging. We attribute it to the limited self-correcting ability of LLMs, as pointed out in \citet{huang2023large}. The feedback mechanisms in Self-Debugging (self-explaining and self-tracing) in fact enhance the initial understanding of coding tasks but fail to align the task to the specific code. Therefore, the debugging performance of these methods quickly converges and then hardly improves even given more rounds of debugging.
In the contrast, \tool exhibits a continuing improvement with the growth of debugging iterations. The new information from runtime execution keeps moving the models towards correct programs, which closely resembles the human debugging process.

\subsection{Different Decomposition Levels of \tool}

In Profiling (Section \ref{subsec:profiling}), we segment the runtime trace into basic blocks based on the control flow graph. A basic block only has one entry and one exit in the program execution, serving as an ideal basic unit in the runtime analysis~\cite{sherwood2001basic}. To explore the effectiveness and efficiency of block-level decomposition, we develop two comparative methods, \tool (line-level) and \tool (function-level), which segment the runtime trace in the granularity of lines and functions respectively. We denote the original block-level debugging method as \tool (block-level). \tool in different decomposition levels share the same architecture. They collect runtime intermediate states at the end of each \emph{code unit}. The code unit can be a line in \tool (line-level), a block in \tool (block-level), or a function in \tool (function-level). Similarly, we adopt Selective Debugging (Section \ref{subsec:debugging}) to fit prompts into the token limits of LLMs. Please refer to Appendix \ref{app:level} for implementation details.

\begin{figure}[t]
    \centering
    \includegraphics[width=0.95\linewidth]{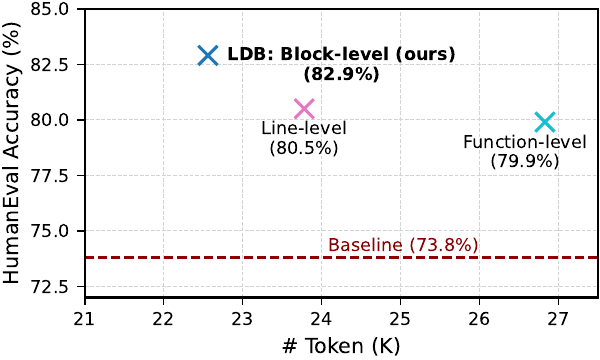}
    
    \caption{Performance vs. average token cost per program of \tool in different decomposition levels on HumanEval with GPT-3.5. \tool (block-level) achieves the best accuracy with the least token cost compared to \tool (line-level) and \tool (function-level).}
    
    \label{fig:level}
\end{figure}

Figure~\ref{fig:level} plots the performance and average token cost per program of \tool in different decomposition levels on HumanEval using GPT-3.5. The detailed statistics are listed in Table~\ref{tab:turn-level} in Appendix. All three debugging methods manage to enhance the performance, demonstrating the benefits from runtime execution information. 
Particularly, among the three decomposition levels, \tool (block-level) achieves the highest improvement. 

\tool (line-level) performs worse than \tool (block-level) even if the line-level information is more fine-grained. This may arise because line-level decomposition leads to incomplete semantics in each code unit (i.e. a line of code). As a result, LLMs struggles to fully understand the code units and accurately identify bugs within the program. 

\tool (function-level) provides the most coarse-grained information which largely preserves the complete programs. However, the intermediate states in the function-level fail to provide detailed runtime information. Consequently, \tool (function-level) is less effective and requires approximately 8.1 iterations on average to debug a program which are much more than the other levels (see Appendix~\ref{app:level} for statistics). This explains the highest token cost of \tool (function-level).




\subsection{Performance of Different Difficulty Levels}

\begin{figure}[t]
    \centering
    \includegraphics[width=0.95\linewidth]{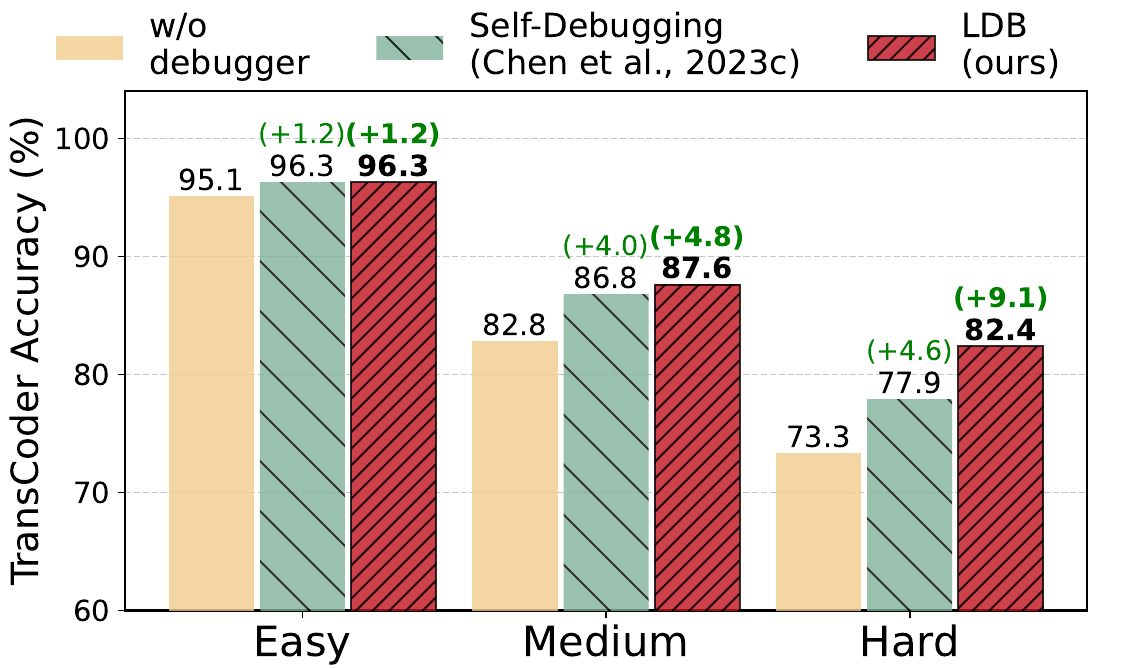}
    
    \caption{Performance of \tool on problems of different difficulty levels in TransCoder using GPT-3.5 as the backbone. \tool demonstrates the most  improvement on Hard-level problems, indicating its capability in understanding the program execution and detecting non-trivial errors in program debugging.}
    
    \label{fig:difficulty}
\end{figure}

To evaluate the capability of \tool in debugging programs, we dive into the problems it successfully debugs in TransCoder and categorize them into three difficulty levels, \emph{Easy}, \emph{Medium}, and \emph{Hard}. The difficulty annotation is automatically performed by GPT-4 (\texttt{1106-preview}) based on the canonical solutions of each problem to avoid potential subjectiveness.
Figure~\ref{fig:difficulty} shows the improved accuracy for each difficulty level from TransCoder dataset. 
We observe that the performance of GPT-3.5 in code generation decreases with the increasing problem difficulty while the improvement from the debugger increases in harder problems. Particularly, \tool shows the most improvement (9.1\%) on the hard-level problems, which indicates that \tool is able to detect the non-trivial bugs and understand the complex execution flows in the harder problems. 

\begin{figure*}[t]
    \centering
    \includegraphics[width=0.98\linewidth]{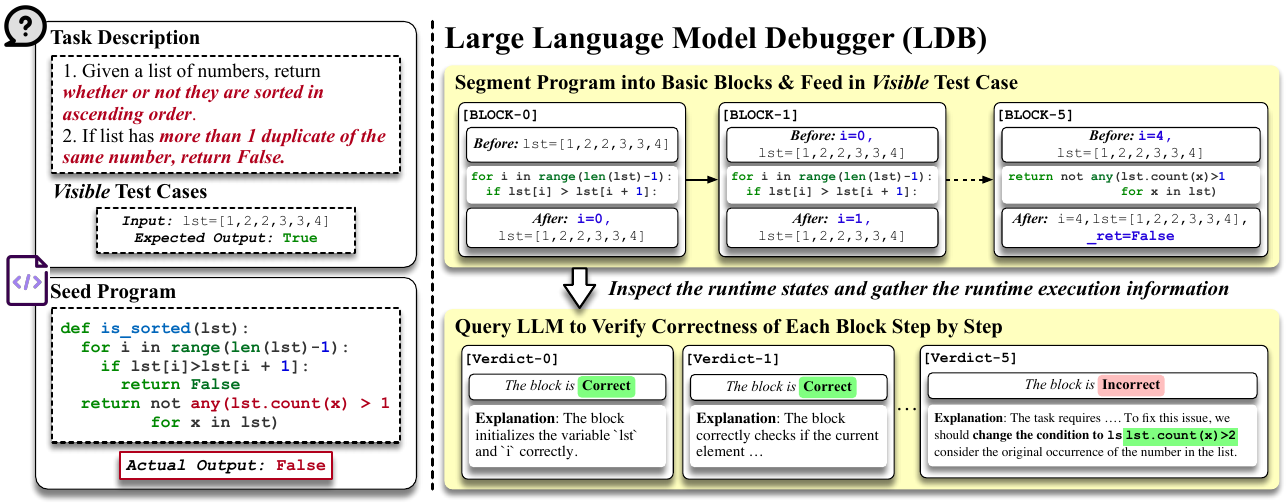}
    
    \caption{Debugging example of \tool on HumanEval with GPT-3.5. \tool accurately detects the bug in Block-5, and suggests tha correct fix to align the program to the task description.}
    
    \label{fig:case-study}
\end{figure*}

\subsection{Case Study}

Figure~\ref{fig:case-study} presents an example on HumanEval with GPT-3.5. In the example, \tool successfully fix the the program and enable it to pass the visible and hidden test cases.
This case requires the program to check two conditions: (1) numbers sorted in ascending order. (2) list does not have more than 1 duplicate of the same number. In the seed program, it checks the first condition, while mistakenly presents the second condition as not having any duplicates (\texttt{lst.count(x) > 1}). In BLOCK-0 to BLOCK-4, \tool makes the verdicts that each block correctly checks the first condition. For BLOCK-5, \tool finds out the mistake in the condition and locates the program bugs exactly in this block. It proposes to fix this condition into \texttt{lst.count(x) > 2} so that it aligns with the task description. 

\section{Related Work}
\paragraph{Augmented Code Language Models} Recent language models based on deep neural networks~\cite{achiam2023gpt, roziere2023code, li2023starcoder, nijkamp2022codegen} demonstrate great potential for coding tasks. Despite their impressive capabilities, these models face challenges such as syntax correctness~\cite{jin2023inferfix, zhang2023syntax}, semantic alignment~\cite{ni2023lever, fan2023automated}, code reliability~\cite{zhong2023study}, and conflict merging~\cite{zhang2022using}. To address these challenges, some focus on enhancing initial code generation by leveraging multiple candidates~\cite{shinn2023reflexion, zhou2023language, gu2023llm} or refining solutions based on better test cases~\cite{zhang2023algo} and self-consistency~\cite{chen2022codet, le2023codechain}, while others train verifiers on execution results to predict solution quality~\cite{ni2023lever}. In contrast, \tool enhances code generation without expanding sampling numbers complementing existing methods, while these existing code generation methods could provide better seeds for our debugger.

\para{Feedback-based Code Refinement} Generating correct solutions could require iterative refinement due to model limitations. Interactive methods like using human feedback~\cite{chen2023improving, wu2023fine} are effective but labor-intensive. Alternatively, code refinement techniques~\cite{chen2023teaching, jiang2023selfevolve, hu2024leveraging} based on language models have been proposed, utilizing interpreter outputs, self-generated explanation, and other users' submission in competition-level programming~\cite{zhang2022automated}. Some train additional models for bug fixing~\cite{pearce2023examining, huang2023empirical, gupta2023grace}, while \tool utilizes debugging capabilities of original large language models. \tool follow an similar iterative refinement paradigm as previous works~\cite{madaan2023self, zhou2023solving}. It leverages execution results for debugging inspired by previous works on execution-guided code generation~\cite{chen2018execution, ni2022learning}.
Additionally, agent frameworks using reinforcement learning for coding tasks~\cite{shinn2023reflexion, zhou2023language, hong2023metagpt, rasheed2024codepori, le2022coderl} incorporate feedback from environments to guide actions. While these frameworks excel in searching code generation space, \tool focuses on code refinement and consistently improves performance across various initial code.

\para{Decomposition in Reasoning} Prompting methods suggest that decomposing problems aids large language models in reasoning tasks~\cite{wei2022chain, zhou2022least, lightman2023let, dhuliawala2023chain,wang2024chain, cheng2022binding}. \cite{zhang2022repairing} proposes program trunking to improve the program fix rate. Inspired by this, \tool decomposes programs into blocks, querying language models for debugging verdicts and explanation. \tool further introduces batch debugging to improve the efficiency. 
\section{Conclusion}
We present \tool, a debugging framework that help LLMs refine generated programs with runtime execution information. We empirically show that \tool significantly improves code generation accuracy and achieves state-of-the-art performance in program debugging, by segmenting the programs into basic blocks and tracking the intermediate values. Experiments also reveal its unique paradigm of program debugging by using runtime information. 
\section*{Limitation}
\tool is a program debugging framework using large language models. Therefore, it is subjected to the limitation of existing debugging methods of human developers. The correct test cases are mandatory in \tool so that \tool can execute the program and compare the execution flow against the task description. It remains an open question in future study whether LLMs are able to do self-correct by simply looking at its intermediate execution without knowing whether the result is correct or not (a.k.a. test-case-free debugging).
\section*{Ethic Statements}
This paper focuses on debugging code generated by large language models. The architecture are built upon open-source models and publicly available proprietary models. All the datasets in this paper are available online. We did not hire any human annotators in our experiments. We will release the code and datasets on \url{https://github.com/FloridSleeves/LLMDebugger}. Therefore, we do not anticipate any major ethical concerns.

\section*{Acknowledgement}
Our work is sponsored in part by NSF CAREER Award 2239440, NSF Proto-OKN Award 2333790, as well as generous gifts from Google, Adobe, and Teradata. Any opinions, findings, and conclusions or recommendations expressed herein are those of the authors and should not be interpreted as necessarily representing the views, either expressed or implied, of the U.S. Government. The U.S. Government is authorized to reproduce and distribute reprints for government purposes not withstanding any copyright annotation hereon.

\bibliography{ref}
\bibliographystyle{acl_natbib}

\appendix

\clearpage

\section*{Appendix}
\section{Performance with Better Reasoners}
We explore the relationship between performance and the ability of debugging backbones. With better reasoners as debugging backbones, \tool achieves higher performance, which can be up to 98.2\% on HumanEval with Reflexion as the seed generators and GPT-4o (\texttt{2024-05-13}) as the backbones.

\begin{table}[h]
   \centering
     \small
        \resizebox{1\linewidth}{!}{
        \setlength{\tabcolsep}{1mm}{
         \begin{tabular}{llll}
         \toprule
         \textbf{Reflexion}  & \textbf{+\tool \scriptsize{(GPT-3.5)}} & \textbf{+\tool \scriptsize{(GPT-4)}} & \textbf {+\tool \scriptsize{(GPT-4o)}}\\
         \midrule
         91.5 & 95.1\scriptsize{\textcolor{cadmiumgreen}{{(+3.6)}}} & 96.9\scriptsize{\textcolor{cadmiumgreen}{{(+5.4)}}} & 98.2 \scriptsize{\textcolor{cadmiumgreen}{{(+6.6)}}} \\
         \bottomrule
         \end{tabular}
         }
        }
     \caption{Accuracy of \tool on HumanEval with seed programs from Reflexion shows that \tool performs better with more advanced debugging backbones. We use GPT-3.5 (\texttt{turbo-0613}), GPT-4 (\texttt{1106-preview}), and GPT-4o (\texttt{2024-05-13}) as the debugging backbone. \tool performs better when it is backboned with better reasoners (GPT-4 and GPT-4o).}
     \label{tab:diff_seed}
\end{table}

\section{Programming Language Concepts}
\label{app:pl}
\paragraph{Basic Block.} A basic block is defined as a linear sequence of code containing a single entry point and a single exit point~\cite{allen1970control, aho2015compilers}. Upon executing the first instruction within a basic block, all subsequent instructions are guaranteed to execute exactly once and in sequential order. A sequence of instructions is considered a basic block if it satisfies two conditions: (1) each instruction in the sequence always executes before all subsequent instructions, and (2) there are no intervening instructions between any two instructions in the sequence~\cite{cocke1970global, allen1970control}.

\paragraph{Control Flow Graph.} The control-flow graph (CFG)~\cite{allen1970control, prosser1959applications} serves as a graphical depiction of all potential paths traversed during the execution of a program. Each node within the CFG corresponds to a basic block, with directed edges representing transitions in the control flow. Typically, two special blocks are identified: the entry block, which signifies the initiation of control flow into the graph, and the exit block, where all control flow exits the graph.

\paragraph{Execution Traces.} In this paper, execution traces are control-flow traces of the whole program~\cite{larus1999whole}. An control-flow trace of a program is a sequence of consecutively executed basic blocks within the program. It also corresponds to a path in the control flow graph from the entry block to the exit block~\cite{ball1994optimally, ball1996efficient, ammons1998improving}.

\section{Implementation Details}
\label{app:param}
In the debugging stage of \tool, we generate the debugging verdicts and explanation using greedy decoding with temperature $T=0$ to improve the reproducibility of our experiment. The maximum number of debugging iterations is 10. 
We set the threshold for the number of sampled blocks and input tokens at 10 and 3,097, respectively.

\section{Overhead Breakdown}
The time cost of \tool is comparable to other baseline methods and all debugging steps of \tool are fully automated without any human labor, as shown in Table~\ref{tab:breakdown}. 
We summarize the standard paradigm of iterative refinement methods as follows:
(1) Execute the buggy program with a Python interpreter. 
(2) Query a LLM to generate debugging feedback.
(3) Query a LLM again to regenerate a program based on the feedback. 
\tool, along with the baseline methods Self-Debugging all follows this paradigm. We compare the time cost of \tool and Self-Debugging in each stage on HumanEval with GPT-3.5.
Compared with the baseline method, the additional time cost of \tool comes from the profiling step (Section~\ref{subsec:profiling}). This step is performed by program analysis and only costs 0.09 seconds on average for programs in HumanEval. The overhead turns out to be negligible under the fluctuation of other overhead while contributing to a significant improvement in the performance as shown in Table~\ref{tab:main} and Figure~\ref{fig:rounds}.
\begin{table*}[!h]
  \centering
      \small
\resizebox{1\linewidth}{!}{
    \setlength{\tabcolsep}{2mm}{
    \begin{tabular}{lll}
    \toprule
    \bf Debugging Stages & \bf \tool & \bf Self-Debugging \\
    \midrule
    Execute the buggy program & 0.01s & 0.01s\\
    Query a LLM to generate feedbacks & 0.09s (profiling) + 9.25s (feedback generation) & 9.84s\\
    Query a LLM to regenerate programs & 7.83s & 7.17s\\
    Other minor overhead & 0.05s & 0.01s \\
    \midrule
    Total Time & 17.23s & 17.08s \\
    \bottomrule
    \end{tabular}%
    }
    }
    \caption{Performance breakdown of \tool and Self-Debugging.}
    \label{tab:breakdown}
\end{table*}

\section{Tradeoffs in Debugging in Different Decomposition Levels}
\label{app:level}
We conduct different level debugging following this design: (1) For \tool (line-level) debugging, we collect the intermediate states before and after each line execution. We sample the first 25 lines and last 25 lines when the line number exceeds threshold $N_b = 50$, which is five times of the block-level threshold. We set this number based on the previous research on the average number of instructions in a basic block~\cite{rotenberg1996trace}. (2) For the original \tool (block-level) debugging, we sample the first 5 blocks and the last 5 blocks when the block number exceeds threshold $N_b=10$. (3) For \tool (function-level) debugging, we decompose the program on function-level, namely we only collect the intermediate states at the entry and exit of the solution function. If the function trace exceeds the context length, we sample first 25 lines and last 25 lines to ensure same amount of code trace information with block-level and line-level. These three level of decomposition expand from fine to coarse granularity. 

We show the average token cost per program and debugging turns of \tool with different granularity debugging levels in Table~\ref{tab:level} and Table~\ref{tab:turn-level}. Using GPT-3.5, \tool (line-level) has less token cost than \tool-Function due to less debugging turns. However, using CodeLlama, an open source model, \tool has higher token cost in the line-level debugging than the function-level debugging. The debugging turns of line-level debugging is not significantly lower than function-level debugging, which shows that CodeLlama has worse reasoning ability for line-level debugging even with more runtime execution information. Besides, line-level debugging has higher token costs for each debugging turns. Therefore, it has the highest token cost. Both GPT-3.5 and CodeLlama demonstrate better efficiency and accuracy in the block-level debugging. Based on these observations, we choose block-level debugging in \tool. 
\begin{table}[!h]
  \centering
      \small
\resizebox{1\linewidth}{!}{
    \setlength{\tabcolsep}{0.8mm}{
    \begin{tabular}{lllll}
    \toprule
    \bf HumanEval & \bf GPT-3.5 & \bf Token & \bf CodeLlama & \bf Token \\
    \midrule
    (w/o debugger) & 73.8 & - & 49.4 & -\\
    \tool (line-level) & 80.5 (+6.7) & 24K  & 53.7 (+4.3) & 72K\\
    \tool (block-level) & 82.9 (+9.1) & 23K & 55.5 (+6.1) & 52K\\
    \tool (function-level) & 79.9 (+6.1) & 27K & 53.7 (+4.3) & 54K\\
    \bottomrule
    \end{tabular}%
    }
    }
    \caption{Accuracy vs average token number per problem on HumanEval. For both GPT-3.5 and CodeLlama, \tool{} with block-level debugging achieves the highest accuracy and least token cost.}
    \label{tab:level}
\end{table}
\begin{table}[!h]

  \centering
      \small
  \resizebox{0.99\linewidth}{!}{
    \setlength{\tabcolsep}{0.5mm}{
    \begin{tabular}{lcccc}
    \toprule
    \bf HumanEval & \bf GPT-3.5 & \bf Avg. Turn & \bf CodeLlama & \bf Avg. Turn \\
    \midrule
    (w/o debugger) & 73.8 & - & 49.4 & -\\
    \tool (line-level) & 80.5 & 6.4  & 53.7 & 9.1\\
    \tool (block-level) & 82.9 & 6.2 & 55.5 & 7.8\\
    \tool (function-level) & 79.9 & 8.1 & 53.7 & 9.3\\
    \bottomrule
    \end{tabular}%
    }
    }
    \caption{Accuracy vs debugging turns on HumanEval. For both GPT-3.5 and CodeLlama, \tool{} with block-level debugging achieves the highest accuracy and fewest debugging turns.}
    \label{tab:turn-level}

\end{table}
\begin{figure}[htb]
    \centering
    \includegraphics[width=0.98\linewidth]{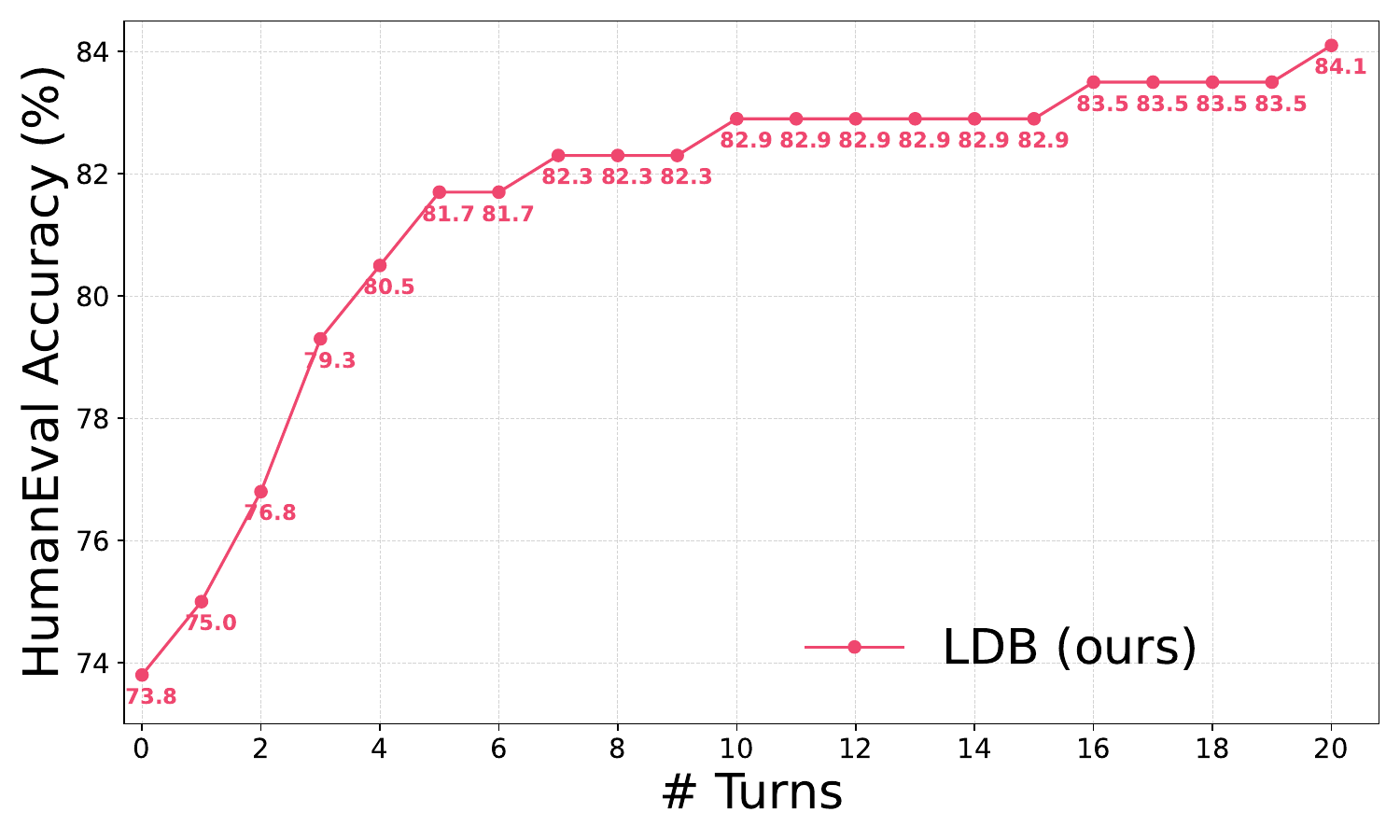}
    
    \caption{Performance of \tool on HumanEval with GPT-3.5 in 20 debugging iterations. The final performance after 20 iterations is 84.1\%.}
    \label{fig:20turns}
\end{figure}

\section{Complexity Analysis of Batch Debugging}
\tool can batch runtime information of all selected blocks together and query language models for debugging verdicts. This significantly improves the token efficiency of \tool and alleviates the pitfall of repeatedly sending context to language models in iterative refinement~\cite{hu2023tree}. Assume the average token numbers is $\mathcal{N}$ for debugging each block, average block number is $\mathcal{B}$ for debugging, and average debugging iteration number is $\mathcal{D}$. Without batch debugging, for debugging $i$-th block, the context length for debugging is $i * \mathcal{N}$. Therefore, debugging $\mathcal{N}$ blocks consumes $\sum_{i=0}^{B}{(i * \mathcal{N})}$, approximately $O(\mathcal{B}^2 * \mathcal{N})$ tokens. As a comparison, batch debugging only send one debugging message in each turn, which has context length of $O(\mathcal{B} * \mathcal{N})$ tokens.

\section{Debugging 10 More Iterations}
As \tool shows a different trend of performance regarding debugging turns compared to the state-of-the-art methods, it is interesting to see whether the performance continues growing after 10 debugging turns. We conduct a experiment that continue debugging until 20 debugging turns to explore the characteristic of \tool, as shown in Figure~\ref{fig:20turns}. The accuracy still grows after 10 debugging turns and achieves 84.1\%.

\section{Error Analysis of \tool}

We analyze the debugging feedback of \tool in two perspectives, the bug localization accuracy and the bug categorization. We evaluate the performance of \tool in localizing the bug accurately and further specify the types of bugs detected by \tool. We use the results with GPT-3.5 as an example.

\paragraph{Accuracy of Bug Localization.} To evaluate the correctness of debugging verdicts made by \tool, we collect the cases that \tool successfully debugs and query GPT-4 (\texttt{1106-Preview}) to automatically verify whether \tool correctly identifies the buggy basic block.
In the prompt, we provide the task description, the seed program, the runtime information, debugging responses, and the final programs for GPT-4.
Table~\ref{tab:correctness} shows the analysis results on HumanEval, MBPP and TransCoder. We observe that \tool achieves an accuracy over 85\% over all datasets. It has the highest accuracy on MBPP (95.5\%) while performs worst on TransCoder (86.7\%). This may arise because TransCoder consists of more complicated programs where the original C++ programs outlines the intention implicitly.

\begin{table}[t]
  \centering
      \small
    \setlength{\tabcolsep}{1mm}{
    \begin{tabular}{lcccccc}
    \toprule
        & \bf HumanEval & \bf MBPP & \bf TransCoder\\
    \midrule
    \bf Accuracy & 93.7\% & 95.3\% & 86.7\% \\
    \midrule
    \bf Syntax Error & 18.8\% & 23.2\% & 20.0\% \\
    \bf Semantic Error & 81.2\% & 76.8\% & 80.0\% \\
    \bottomrule
    \end{tabular}%
    }
    \caption{Debug correctness of \tool{} and detected program error types.}
    \label{tab:correctness}
\end{table}

\paragraph{Bug Categorization.} We further categorize the bugs detected and fixed by \tool into two categories: Syntax Errors and Semantic Errors. Syntax errors refer to the cases that violates the syntax rules of its programming language.
Semantic error occurs when the code runs without producing any syntax errors, but it does not behave as intended, thus fails the test cases. We annotate the categories through GPT-4 (\texttt{1106-preview}) to avoid subjectiveness. Across three datasets, \tool fixes semantic errors with around 80\% of the time. The reason is that the syntax errors are less frequent given the code generation ability of current LLMs, and \tool can better debug semantic errors by introducing the detailed runtime execution information. 

\section{Extendability of \tool}
In this paper, we have chosen Python for \tool due to its widespread popularity in software development~\cite{huang2021pylive}. This aligns with previous studies such as Self-Debugging~\cite{chen2023teaching}, Reflexion~\cite{shinn2023reflexion}, and Gorilla~\cite{patil2023gorilla}, which also utilize Python to showcase the effectiveness of their methodologies. 

We further elaborated on \tool's capability to adapt to other programming languages. Importantly, \tool does not use features that are specific to Python. This makes it easy to adapt \tool for use with other programming languages by changing its program executors and analysis tools~\cite{zhong2023survey, huang2023protecting, zhong2024effective}. In the debugging process of \tool, the backbone LLM debugs programs based on the detailed runtime execution information, such as execution traces and intermediate variables. This runtime execution information can be similarly obtained in all different kinds of programming languages. We can use off-the-shell program analysis tools to instrument the targeted language and acquire the runtime execution information. Incorporating multilingual experiment settings, such as the HumanEval-X, remains as a future work.

\section{Prompt Example in \tool}

We list the prompts used in \tool in detail as follow,
\begin{itemize}[leftmargin=*,nosep]
    \item The prompts used in the chat mode are shown in Figure~\ref{fig:app-demo-chat}.
    \item The prompts used in the text completion mode are shown in Figure~\ref{fig:app-demo-completion}.
\end{itemize}



\clearpage


\begin{figure*}[t]
\onecolumn
\begin{tcolorbox}[left=1mm,right=1mm,top=0.mm, bottom=0mm,colback=white]
\begin{lstlisting}[style=demo,escapechar=!]
==================================== System ====================================
You are an expert programming assistant.
\end{lstlisting}
\end{tcolorbox}
\begin{tcolorbox}[left=1mm,right=1mm,top=0.mm, bottom=0mm,colback=white]
\begin{lstlisting}[style=demo,escapechar=!]
===================================== User =====================================
Complete the following task in Python. Please respond with code only.
def is_sorted(lst):
    '''
    Given a list of numbers, return whether or not they
    are sorted in ascending order. If list has more than 
    1 duplicate of the same number, return False. 
    Assume no negative numbers and only integers.
    
    Examples
    is_sorted([5]) = True
    is_sorted([1, 2, 3, 4, 5]) = True
    is_sorted([1, 2, 3, 4, 5, 6, 7]) = True
    is_sorted([1, 2, 2, 2, 3, 4]) = False
    '''
\end{lstlisting}
\end{tcolorbox}
\begin{tcolorbox}[left=1mm,right=1mm,top=0.mm, bottom=0mm,colback=white]
\begin{lstlisting}[style=demo,escapechar=!]
=================================== Assistant ==================================
    for i in range(len(lst) - 1):
        if lst[i] > lst[i + 1]:
            return False
    return not any(lst.count(x) > 1 for x in lst)
\end{lstlisting}
\end{tcolorbox}
\begin{tcolorbox}[left=1mm,right=1mm,top=0.mm, bottom=0mm,colback=white]
\begin{lstlisting}[style=demo,escapechar=!]
===================================== User =====================================
The code above fails the given unit test:
assert is_sorted([1, 2, 2, 3, 3, 4]) == True # Real Execution Output: False. 
Help me debug this.

Here is the code execution trace block by block with the intermediate variable values to debug your code. You need to answer FOR EACH BLOCK whether this code block is correct or not. If not, give an explanation on what is wrong. Please wrap your response into a JSON object that contains keys `block` with the name of each block, key `correct` with value False or True, and key `explanation` with an explanation on the bug. 

Example Answers:
{"block": "BLOCK-1", "correct": "True", "explanation": "The block initializes variable `a` and `b`."}
{"block": "BLOCK-2", "correct": "False", "explanation": "The block is incorrect because the code does not add the two integers together, but instead subtracts the second integer from the first. To fix this issue, we should change the operator from `-` to `+` in the return statement. This will ensure that the function returns the correct output for the given input."}

[BLOCK-0]
    # lst=[1, 2, 2, 3, 3, 4]
    for i in range(len(lst) - 1):
        if lst[i] > lst[i + 1]:
    # i=0	lst=[1, 2, 2, 3, 3, 4]
[BLOCK-1]
    # i=0	lst=[1, 2, 2, 3, 3, 4]
    for i in range(len(lst) - 1):
        if lst[i] > lst[i + 1]:
    # i=1	lst=[1, 2, 2, 3, 3, 4]
[BLOCK-2]
    # i=1	lst=[1, 2, 2, 3, 3, 4]
    for i in range(len(lst) - 1):
        if lst[i] > lst[i + 1]:
    # i=2	lst=[1, 2, 2, 3, 3, 4]
[BLOCK-3]
    # i=2	lst=[1, 2, 2, 3, 3, 4]
    for i in range(len(lst) - 1):
        if lst[i] > lst[i + 1]:
    # i=3	lst=[1, 2, 2, 3, 3, 4]
\end{lstlisting}
\end{tcolorbox}
\end{figure*}
\begin{figure*}[t]
\begin{tcolorbox}[left=1mm,right=1mm,top=0.mm, bottom=0mm,colback=white]
\begin{lstlisting}[style=demo,escapechar=!]
[BLOCK-4]
    # i=3	lst=[1, 2, 2, 3, 3, 4]
    for i in range(len(lst) - 1):
        if lst[i] > lst[i + 1]:
    # i=4	lst=[1, 2, 2, 3, 3, 4]
[BLOCK-5]
    # i=4	lst=[1, 2, 2, 3, 3, 4]
    for i in range(len(lst) - 1):
    return not any(lst.count(x) > 1 for x in lst)
    # i=4	lst=[1, 2, 2, 3, 3, 4]	_ret=False
\end{lstlisting}
\end{tcolorbox}
\begin{tcolorbox}[left=1mm,right=1mm,top=0.mm, bottom=0mm,colback=white]
\begin{lstlisting}[style=demo,escapechar=!]
=================================== Assistant ==================================
{"block": "BLOCK-0", "correct": true, "explanation": "The block initializes the variable `lst` correctly."}
{"block": "BLOCK-1", "correct": true, "explanation": "The block correctly checks if the current element is greater than the next element in the list."}
{"block": "BLOCK-2", "correct": true, "explanation": "The block correctly checks if the current element is greater than the next element in the list."}
{"block": "BLOCK-3", "correct": true, "explanation": "The block correctly checks if the current element is greater than the next element in the list."}
{"block": "BLOCK-4", "correct": true, "explanation": "The block correctly checks if the current element is greater than the next element in the list."}
{"block": "BLOCK-5", "correct": false, "explanation": "The block is incorrect because it returns the opposite of the condition `lst.count(x) > 1` for any element `x` in the list. This means that if any element has more than 1 duplicate, the function will return False. However, the task requires that if there are more than 1 duplicate of the same number, the function should return False. To fix this issue, we should change the condition to `lst.count(x) > 2` to account for the original occurrence of the number in the list."}
\end{lstlisting}
\end{tcolorbox}
\begin{tcolorbox}[left=1mm,right=1mm,top=0.mm, bottom=0mm,colback=white]
\begin{lstlisting}[style=demo,escapechar=!]
===================================== User =====================================
Please fix the Python code.
\end{lstlisting}
\end{tcolorbox}
\begin{tcolorbox}[left=1mm,right=1mm,top=0.mm, bottom=0mm,colback=white]
\begin{lstlisting}[style=demo,escapechar=!]
=================================== Assistant ==================================
def is_sorted(lst):
    '''
    Given a list of numbers, return whether or not they
    are sorted in ascending order. If list has more than 
    1 duplicate of the same number, return False. 
    Assume no negative numbers and only integers.
    
    Examples
    is_sorted([5]) = True
    is_sorted([1, 2, 3, 4, 5]) = True
    is_sorted([1, 2, 3, 4, 5, 6, 7]) = True
    is_sorted([1, 2, 2, 2, 3, 4]) = False
    '''
    for i in range(len(lst) - 1):
        if lst[i] > lst[i + 1]:
            return False
    return not any(lst.count(x) > 2 for x in lst)
\end{lstlisting}
\end{tcolorbox}
\caption{Prompt for chat mode.}
\label{fig:app-demo-chat}
\end{figure*}

\clearpage


\begin{figure*}[t]
\begin{tcolorbox}[left=1mm,right=1mm,top=0.mm, bottom=0mm,colback=white]
\begin{lstlisting}[style=demo,escapechar=!]
# Write Python function to complete the task and pass the assertion tests.
### Task Start ###
# These are the assertions for your function:
assert find_char_long('Please move back to stream') == ['Please', 'move', 'back', 'stream']

def find_char_long(text):
    """ 
    Write a function to find all words which are at least 4 characters long in a string by using regex. 
    """
    if text == "":
        return []
    pat = r"\b\w{4}\b"
    res = re.findall(pat, text)
    return res

With the above function, the assertion is `find_char_long('Please move back to stream') == ['Please', 'move', 'back', 'stream']` but the real execution output is `['move', 'back']`.

Debug the program trace block by block until find the incorrect block. Every block should have different feedbacks:
[BLOCK-0]
    # text="Please move back to stream"
    if text == "":
[BLOCK-1]
    # text="Please move back to stream"
    pat = r"\b\w{4}\b"
    res = re.findall(pat, text)
    # text="Please move back to stream" pat="\b\w{4}\b"  res=['move', 'back']
[debug]
[BLOCK-0]
Feedback: CORRECT. This block is correct. It checks if the input text is empty. If the input text is empty, it returns an empty list without do regex match.
[BLOCK-1]
Feedback: INCORRECT. This block defines a regular expression pattern `pat` with value r"\b\w{4}\b". However, there's an issue with the regular expression pattern. It only matches words that are exactly 4 characters long. Therefore, the return value `_ret` is `['move', 'back']`. In the task description, it asks for words which are at least 4 characters long. To fix the code, we should change`pat=r"\b\w{4}\b"` into `pat = r"\b\w{4,}\b"`.
[/debug]
Please fix the Python code.
[python]
import re
def find_char_long(text):
    """ 
    Write a function to find all words which are at least 4 characters long in a string by using regex. 
    """
    if text == "":
        return []
    pat = r"\b\w{4,}\b"
    res = re.findall(pat, text)
    return res
[/python]
### Task End ###

### Task Start ###
# These are the assertions for your function:
assert solution([3, 3, 3, 3, 3]) == 9 

def solution(lst):
    """
    Given a non-empty list of integers, return the sum of all of the odd elements that are in even positions.
\end{lstlisting}
\end{tcolorbox}
\end{figure*}

\begin{figure*}[t]
\begin{tcolorbox}[left=1mm,right=1mm,top=0.mm, bottom=0mm,colback=white]
\begin{lstlisting}[style=demo,escapechar=!]
    Examples
    solution([5, 8, 7, 1]) ==> 12
    solution([3, 3, 3, 3, 3]) ==> 9
    solution([30, 13, 24, 321]) ==>0
    """

    result = 0
    for i in range(len(lst)):
        if lst[i] % 2 == 0 and i % 2 == 0:
            result += lst[i]
    return result

With the above function, the assertion is `assert solution([3, 3, 3, 3, 3]) == 9 ` but the real execution output is ` 0`.

Debug the program trace block by block until find the incorrect block. Every block should have different feedbacks:
[BLOCK-0]
    # lst=[3, 3, 3, 3, 3]
    result = 0
    for i in range(len(lst)):
        if lst[i] % 2 == 0 and i % 2 == 0:
    # lst=[3, 3, 3, 3, 3] result=0 i=0
[BLOCK-1]
    # lst=[3, 3, 3, 3, 3] result=0 i=0
    for i in range(len(lst)):
        if lst[i] % 2 == 0 and i % 2 == 0:
    # lst=[3, 3, 3, 3, 3] result=0 i=1
[BLOCK-2]
    # lst=[3, 3, 3, 3, 3] result=0 i=1
    for i in range(len(lst)):
        if lst[i] % 2 == 0 and i % 2 == 0:
    # lst=[3, 3, 3, 3, 3] result=0 i=2
[BLOCK-3]
    # lst=[3, 3, 3, 3, 3] result=0 i=2
    for i in range(len(lst)):
        if lst[i] % 2 == 0 and i % 2 == 0:
    # lst=[3, 3, 3, 3, 3] result=0 i=3
[BLOCK-4]
    # lst=[3, 3, 3, 3, 3] result=0 i=3
    for i in range(len(lst)):
        if lst[i] % 2 == 0 and i % 2 == 0:
    # lst=[3, 3, 3, 3, 3] result=0 i=4
[BLOCK-5]
    # lst=[3, 3, 3, 3, 3] result=0 i=4
    for i in range(len(lst)):
    return result
    # lst=[3, 3, 3, 3, 3] result=0 i=4 _ret=0
[debug]
\end{lstlisting}
\end{tcolorbox}
\end{figure*}
\begin{figure*}[t]
\begin{tcolorbox}[left=1mm,right=1mm,top=0.mm, bottom=0mm,colback=white]
\begin{lstlisting}[style=demo]
=================================== Completion ==================================
[BLOCK-0]
Feedback: CORRECT. This block is correct. It initializes the result variable to 0. 
[BLOCK-1]
Feedback: INCORRECT. This block has an issue. It checks if the element at index `i` is even and if the index `i` is even. However, the task description asks for the sum of all of the odd elements that are in even positions. To fix the code, we should change the line `if lst[i] % 2 == 0 and i % 2 == 0:` into `if lst[i] % 2 != 0 and i % 2 == 0:`.
[BLOCK-2]
Feedback: INCORRECT. This block has an issue. It checks if the element at index `i` is even and if the index `i` is even. However, the task description asks for the sum of all of the odd elements that are in even positions. To fix the code, we should change the line `if lst[i] % 2 == 0 and i % 2 == 0:` into `if lst[i] % 2 != 0 and i % 2 == 0:`.
[BLOCK-3]
Feedback: INCORRECT. This block has an issue. It checks if the element at index `i` is even and if the index `i` is even. However, the task description asks for the sum of all of the odd elements that are in even positions. To fix the code, we should change the line `if lst[i] % 2 == 0 and i % 2 == 0:` into `if lst[i] % 2 != 0 and i % 2 == 0:`.
[BLOCK-4]
Feedback: INCORRECT. This block has an issue. It checks if the element at index `i` is even and if the index `i` is even. However, the task description asks for the sum of all of the odd elements that are in even positions. To fix the code, we should change the line `if lst[i] % 2 == 0 and i % 2 == 0:` into `if lst[i] % 2 != 0 and i % 2 == 0:`.
[BLOCK-5]
Feedback: CORRECT. This block is correct. It returns the result variable.
[/debug]
Please fix the Python code.
[python]
def solution(lst):
    """
    Given a non-empty list of integers, return the sum of all of the odd elements that are in even positions.
    
    Examples
    solution([5, 8, 7, 1]) ==> 12
    solution([3, 3, 3, 3, 3]) ==> 9
    solution([30, 13, 24, 321]) ==>0
    """
    result = 0
    for i in range(len(lst)):
        if lst[i] % 2 != 0 and i % 2 == 0:
            result += lst[i]
    return result
[/python]
### Task End ###
\end{lstlisting}
\end{tcolorbox}
\caption{Prompt for completion mode.}
\label{fig:app-demo-completion}
\end{figure*}





\end{document}